\documentstyle[12pt, aasms4]{article}

\lefthead{Bloom et~al.}
\righthead{Variability in Cygnus X-1}

\begin{document}

\title{A Search for Aperiodic Millisecond Variability in Cygnus X-1}

\author{C.~Chaput\altaffilmark{1}, E.~Bloom\altaffilmark{1},
L.~Cominsky\altaffilmark{1, 2}, G.~Godfrey\altaffilmark{1},
P.~Hertz\altaffilmark{3},
J.~Scargle\altaffilmark{4}, G.~Shabad\altaffilmark{1},
H.~Wen\altaffilmark{1}, K.~Wood\altaffilmark{3} and
D.~Yentis\altaffilmark{3}}

\altaffiltext{1}{Stanford Linear Accelerator Center}
\altaffiltext{2}{Sonoma State University}
\altaffiltext{3}{Naval Research Laboratory}
\altaffiltext{4}{NASA Ames Research Center}

\begin{abstract}
We have conducted a search for aperiodic millisecond variability in the 
integrated 1 to 25~keV X-ray region
of Cyg~X-1.  We have examined HEAO~A-1 archival data and 
Rossi X-ray Timing Explorer
(RXTE) guest
observer data for evidence of excess power above the Poisson noise
floor using the relative
integral power analysis and the Fourier transform method. 
Our results are in disagreement with the
results of Meekins \it et~al. \rm(1984). We attribute
the discrepancy to an instrumental effect for which Meekins \it et~al. \rm (1984)
did not apply a correction.  With the correction we see no evidence
for excess power above 25~Hz in the HEAO A-1 data.  
Our analysis of RXTE data is
in agreement with previously published results (\cite{cui97,bell96})
of different data sets and shows no sign of excess power above 30 Hz.

\end{abstract}

\section{Introduction}

Identifying and understanding short timescale variability in cosmic
sources has repeatedly led to a better understanding of their fundamental
nature and of the important physical processes present. For example,
during the last decade, studies of fast time variability in low-mass
X-ray binaries concentrated on quasi-periodic oscillations
(QPOs) and various noise components
frequencies up to 2~kHz. The study of QPOs and associated noise
components led to a qualitatively more complete understanding of the
accretion processes and the various omnipresent instabilities
(\cite{vdk97}). Similar breakthroughs in our understanding
of black hole candidates have yet to occur.

The dynamical and radiative timescales in the inner disk of accreting
black hole candidates are predicted to be in the millisecond
range. The thin
disk models of Wallinder, Kato \& Abramowicz
(1992), scaled to stellar mass black
holes, have local thermal and acoustic timescales $<$ 1 ms, and
quasiperiodic variability in the X-ray emission is predicted at these
timescales. Bao \& $\O$stgaard (1995) have numerically modeled orbiting
shots in a geometrically thin accretion disk around a black hole including
all relativistic effects. The shots, or ``hot spots'', were simulated to
radiate photons isotropically in their proper rest frames. For various
shot distributions and different inclination angles, they find that the
power density spectrum (PDS)
exhibits a ``cutoff'' at the Keplerian frequency corresponding to the
inner edge of the accretion disk. This cutoff is present for both
optically thick and thin disks. The accretion model of
Nowak \& Wagoner (1995) also predicts a sharp cutoff in the
PDS falling
as $f^{-5}$ for frequencies greater than the Keplerian frequency of the
inner edge of the disk. The $f^{-5}$ dependence arises from the
three-dimensional hydrodynamic turbulent flow interior to the edge of
the disk.

Advection-dominated disk models for black holes
(\cite{chak96,nara96}) provide a
self-consistent explanation for the energy spectra of hard and soft
states of black hole candidate binary systems. In these types of
models a standing shock can develop in the accretion flow at about
10 --- 30 Schwarzschild radii
(R$_{\rm{Sch}}$). The location of the shock defines an effective inner
edge for both the disk and the halo components that can lead to
abrupt changes in the PDS. Depending on the mass and angular momentum
of the black hole these effects are predicted to be in the 3 --- 100 Hz
range. There are also interesting theoretical predictions of QPOs at a few
hundred Hz arising from the special character of BH accretion
(\cite{perez97}).

Millisecond variability in Cyg~X-1 has been reported twice.
Rothschild \it et~al. \rm(1974) reported millisecond bursts in an
observation of Cyg~X-1 obtained with a rocket experiment. These
bursts appeared as excess counts over that expected from Poisson
statistics assuming that the Poisson expectation remains constant.
However, the leakage of variability at lower frequencies ($\sim10$
Hz) into the higher frequencies of interest ($\sim1000$ Hz)
invalidates this assumption (\cite{press74,weiss78}). Indeed, when
the pre-1978 literature is carefully reviewed, the analysis of Cyg
X-1 timing spectra from a number of experiments show no conclusive
evidence for millisecond variability (\cite{weiss78}). More recent
results show no model-independent evidence for millisecond
variability (\cite{loch89}), except in the context of the shot
model (\cite{loch91,nego95}). Lochner \it et~al. \rm(1991) used
the phase portrait idea to determine parameters of a shot noise
model.  Using data from HEAO A-2 and EXOSAT,
they find evidence for characteristic shot durations lasting from
milliseconds to a few seconds. However, this analysis is 
model-dependent.

Meekins \it et~al. \rm (1984) (hereafter M84 ) worked to untangle
leakage effects from slower time scales and developed a  $\chi^{2}$
method to claim detected variability at timescales of 0.3 -- 3.0
millisecond in a HEAO A-1 observation of Cyg~X-1 with 8 $\mu$s resolution.
This result has been prominently quoted as the only strong
evidence for variability in Cyg~X-1 at millisecond time scales; for example,
see van der Klis (1995) and Liang (1998). In this paper we present
a reanalysis of HEAO A-1 data and an analysis of 
Rossi X-ray Timing Explorer 
(RXTE) data that
contradicts the apparently clear observation in the PDS of
millisecond power from Cyg~X-1 of M84. From our analyses, we
conclude that the observed millisecond power in the M84 PDS is due
to either the known HEAO A-1 reset
problem (\cite{wood84}) or a previously unknown instrumental effect.

\section{Analyses}

\subsection{Observations}
We have analyzed archival observations of Cyg~X-1 and the
supernova remnant Cas A from the High Energy Astrophysics
Observatory A-1 experiment (HEAO A-1). We have also analyzed new
observations of Cyg~X-1 made by the Rossi X-ray Timing Explorer
(RXTE). The HEAO A-1 observations were made on 1978 May 7 while
Cyg~X-1 was in the hard (low) state.  Cas A was observed on 1978 August
2. Data from this presumably Poisson source were used to model the
response of the detector and to search for 
instrumental effects. Both sets of A-1 data were recorded using
the high bit rate mode described below. The RXTE observations
occurred on 1996 June 8, 1996 June 17, 1996 June 27 and 1996 July
12 as part of our approved RXTE Guest Investigator program.
Similar observations have been previously published (\cite{cui97,bell96}). 
Cyg~X-1 was in its soft (high) state at the time of the
RXTE observations. 
Table~\ref{observ.tbl} shows the observation
dates and times for all of the data used in our analyses.

\placetable{observ.tbl}

We have used two techniques to analyze the HEAO A-1 data: the
relative integral power method of M84 and the standard Fast
Fourier transform (FFT) power spectrum method.  The RXTE
observations were analyzed using the FFT power spectrum method
only.  For the HEAO A-1 analyses, we have derived new methods to
correct the data for both dead time and instrumental effects. For
the RXTE analysis we use the standard RXTE dead time correction
(\cite{zhang95}).

\subsection{Analysis of the HEAO A-1 Data of Cas A and Cyg~X-1}

The HEAO A-1 data were recorded in the high bit rate (HBR) mode
and consisted of a series of zeros and ones.  A zero indicated no
photons in the previous 8 $\mu$s and a one indicated that at least
one photon was detected in the 8 $\mu$s interval.  There was no
energy information in this mode.  The energy range covered by
these observations is about 1 to 25 keV.

\begin{figure}
\plotone{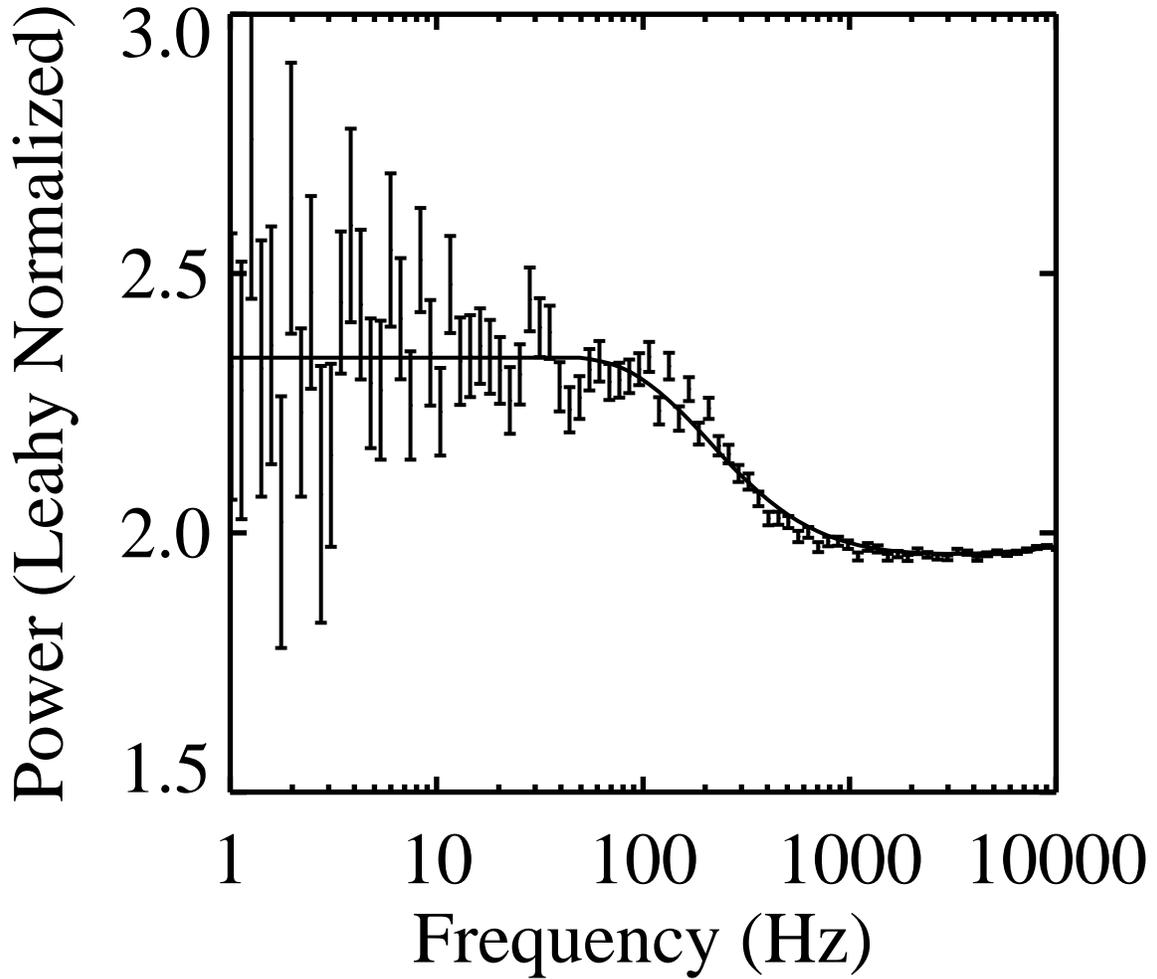} \caption{The HEAO A-1 Leahy-normalized
Power Density Spectrum (PDS) for Cas A. The solid line represents
the best fit PDS assuming the underlying difference in photon 
arrival times ($\Delta t_{\gamma}$)
distribution is given by Equation~\ref{hyper}.} \label{casapwr.eps}
\end{figure}

\begin{figure}
\plotone{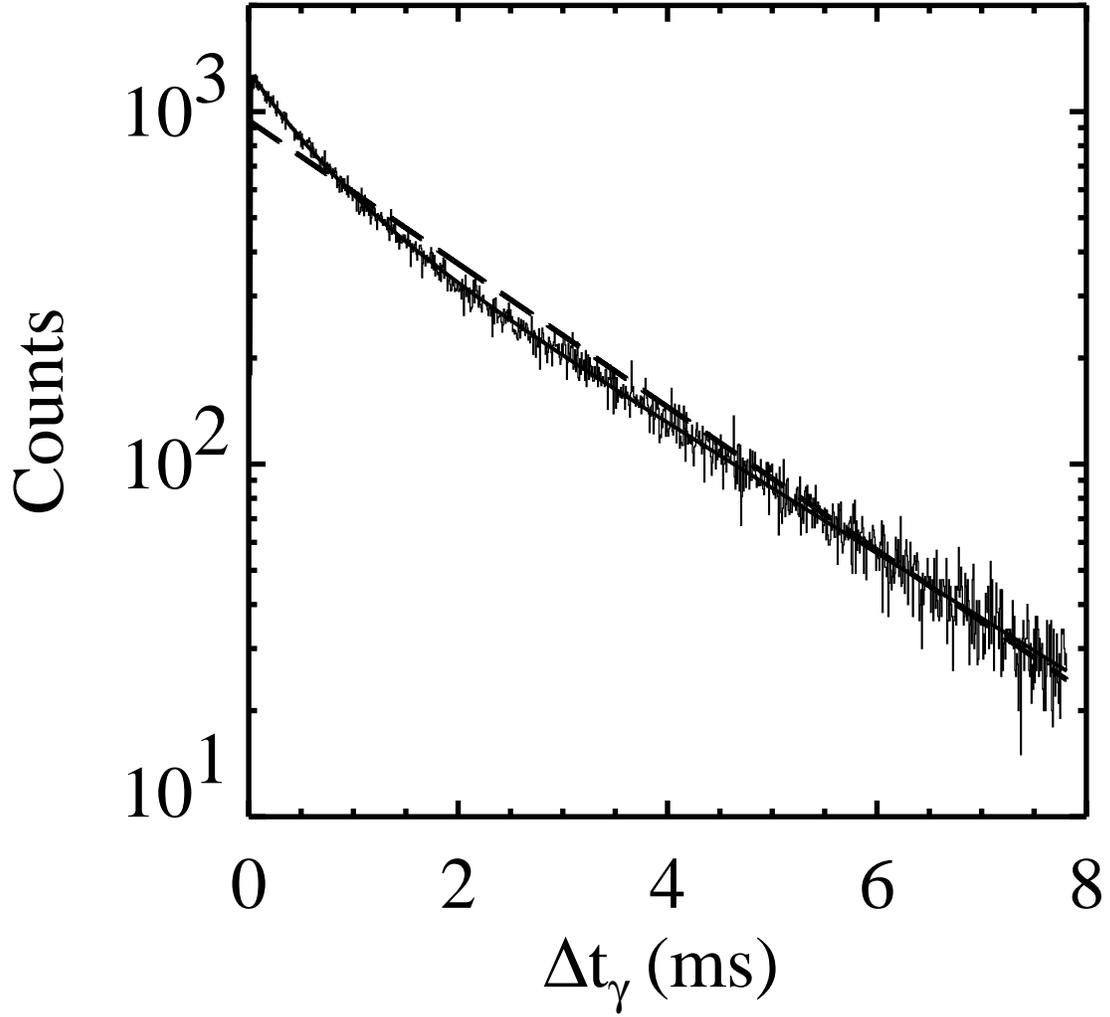}
\caption{The HEAO A-1 difference in
photon arrival times ($\Delta t_{\gamma}$)
distribution for Cas A.  The solid line
represents the best fit of Equation~\ref{hyper} to the distribution.  The
dashed line represents the best fit of a simple offset exponential to the
distribution.}
\label{casawt.eps}
\end{figure}

The data for Cas A were Fourier transformed to search for
deviations from the expected Poisson source spectrum.  For a
Poisson source measured by a detector with no dead time, the
expected spectrum is flat with a value 2, using the Leahy
normalization (\cite{leahy}). 
Introducing dead time into the system slightly reduces the
normalization value, but the shape remains relatively flat
in the region in which we are interested. The observed Fourier
power for Cas A-1 was not flat, but instead showed a broad
``knee'' in the spectrum, as shown in Figure~\ref{casapwr.eps}.
\placefigure{casapwr.eps} The distribution of
differences in photon arrival
times ($\Delta t_{\gamma}$)
showed a kink in the expected offset exponential distribution.
This effect was modeled under the assumption that it
was a previously uncorrected instrumental effect.  
The effect may have been unique to the HBR data or general to 
the HEAO A-1 data, but it would have been difficult to 
observe in the well-studied 5 and 320 ms binned data modes.
In the HBR PDS it was apparent
only at frequencies above about 100 Hz, which is the Nyquist
frequency for the 5 ms data.  It was not possible to
determine the times between individual events for the 5 and 320 ms
modes since the data were binned; therefore the kink in the
$\Delta t_{\gamma}$ distribution is not likely to be observable.  
We searched 
the 5 ms data for this effect and did not find any indications of it.
Eadie \it et al. \rm
(1971) note that the hyperexponential function is applicable in
situations where there is a mixture of exponential processes.  We
find that an offset hyperexponential function is a good representation
of the HEAO A-1 HBR $\Delta t_{\gamma}$ distribution:
\begin{equation}
f_{H}(t)=U(t-\tau)(p_{1}\rho_{1}e^{-\rho_{1}(t-\tau)}+
(1-p_{1})\rho_{2}e^{-\rho_{2}(t-\tau)})
\label{hyper}
\end{equation}
where $U(t-\tau)$ is the Heavyside step function, $\tau$ is the
dead time, $\rho_{1}$ and $\rho_{2}$ are the count rates for two
Poisson processes and $p_{1}$ is the probability of generating an
$\Delta t_{\gamma}$ from the first Poisson process.  The model was fitted
to the Cas A data and the results are shown in
Figure~\ref{casawt.eps}. \placefigure{casawt.eps}

\subsection{Power Spectrum Analysis of HEAO A-1 Data}

\begin{figure}
\plotone{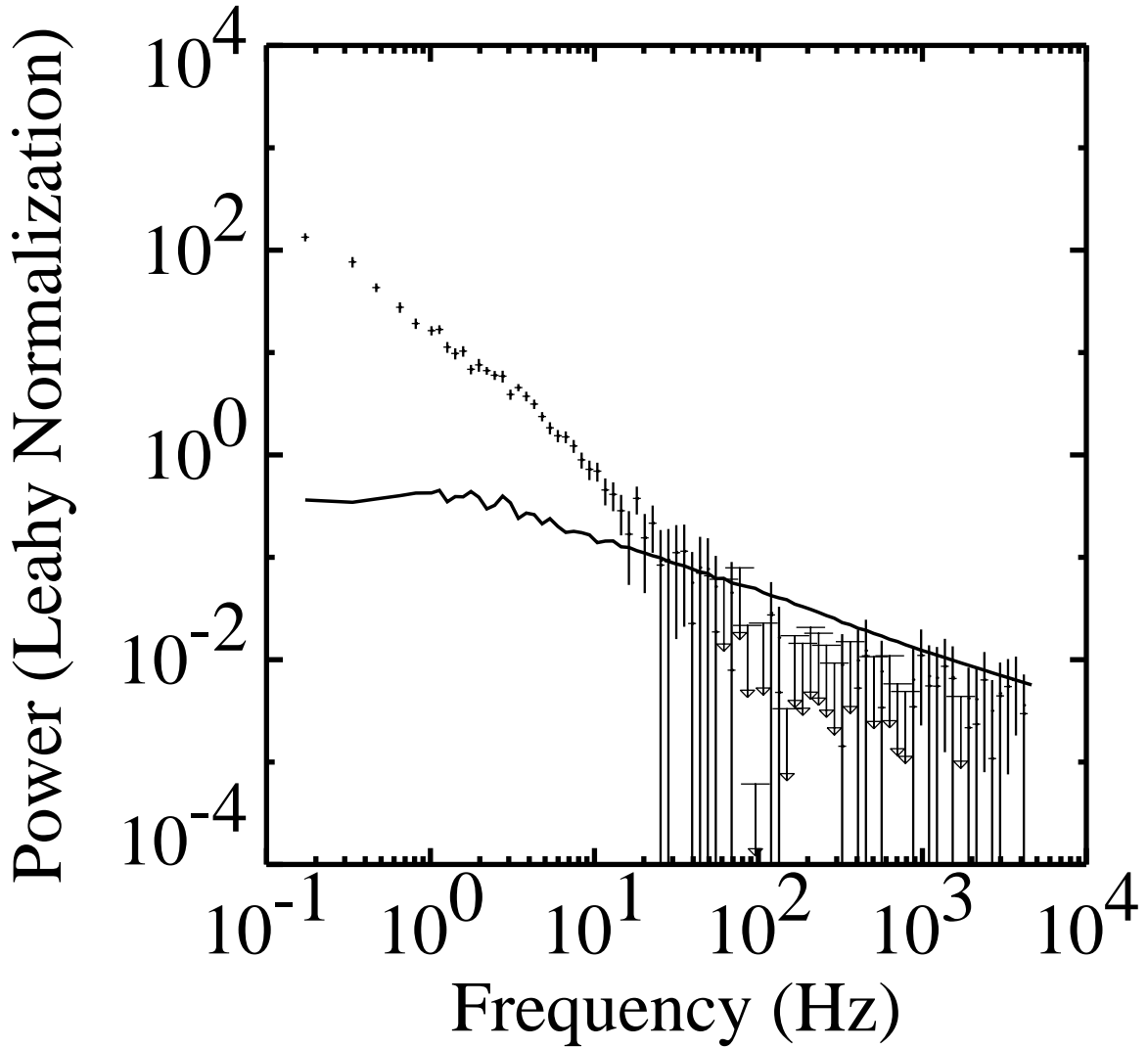} \caption{The HEAO A-1 Leahy-normalized
noise subtracted PDS for Cyg~X-1, using Equation~\ref{hyper} as the
underlying difference in
photon arrival times ($\Delta t_{\gamma}$)
distribution.  The solid line
represents the 95\% confidence level upper limit for detecting
excess power above the Poisson noise floor.} \label{hcygpwr2.eps}
\end{figure}

\begin{figure}
\plotone{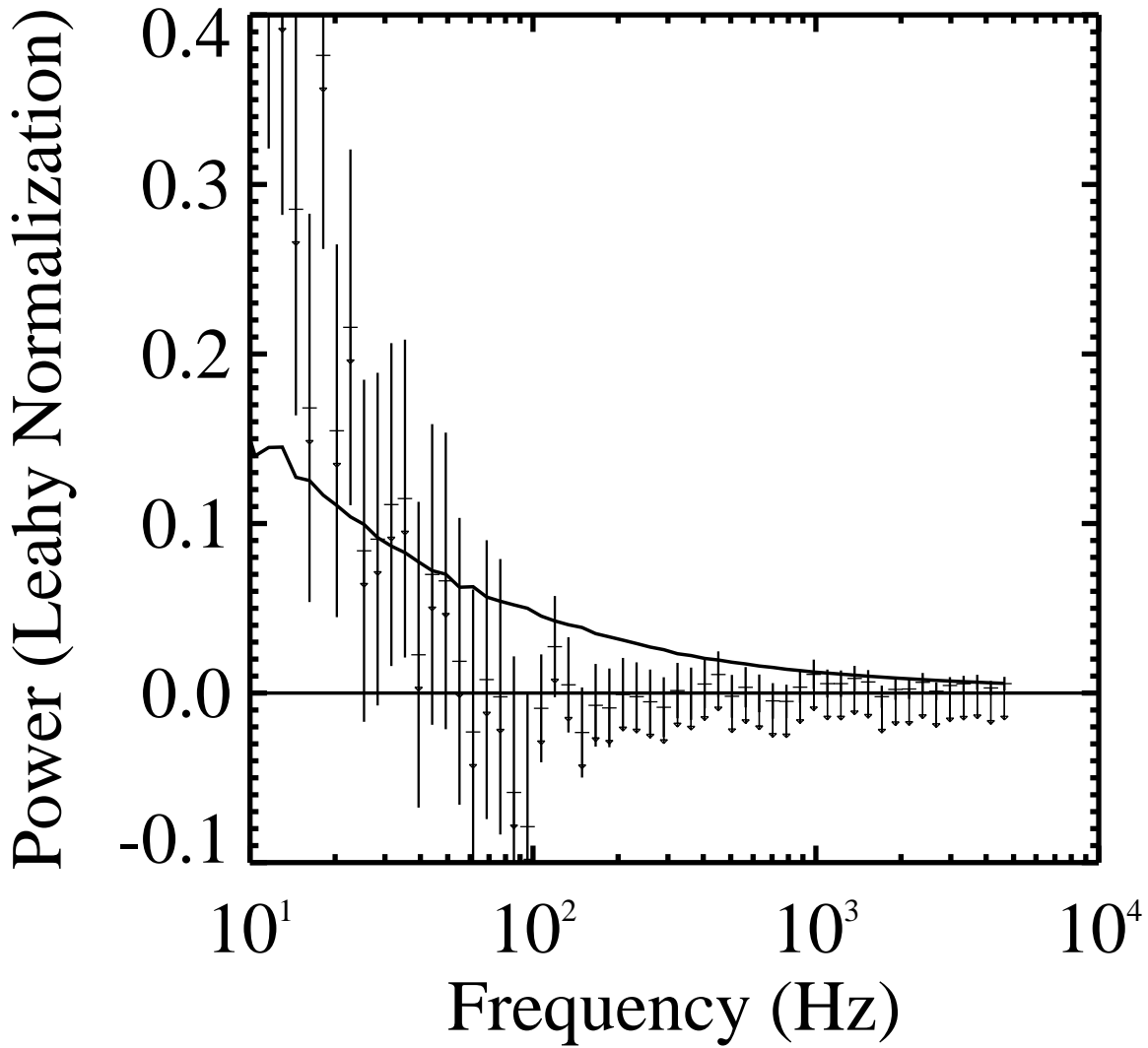} \caption{The HEAO A-1 Leahy-normalized
noise subtracted PDS for Cyg~X-1, using Equation~\ref{hyper} for the
underlying difference in 
photon arrival times ($\Delta t_{\gamma}$) distribution, 
and expanded to show
the high frequency region.  The upper solid line represents the 95\%
confidence level upper limit for detecting excess power above the
Poisson noise floor, and is consistent with zero excess power for
frequencies greater than 40~Hz.  The lower solid line represents
the zero excess power line.} \label{hcygpwr3.eps}
\end{figure}

We analyzed the Cyg~X-1 data using a Fourier transform and
observed a spectrum similarly shaped to the Cas A PDS.  We again
interpreted this as a manifestation of either the instrument reset
problem or of a previously unreported instrumental effect. 
We determined the effective Poisson noise floor in the
presence of the instrumental effect for a non-Poisson source,
Cyg~X-1.  Using the following procedure we fit Equation~\ref{hyper}
to the Cyg~X-1 $\Delta t_{\gamma}$ distribution.  Random 
$\Delta t_{\gamma}$ 
were drawn from Equation~\ref{hyper} as defined by the above 
fit parameters and accumulated to generate absolute times.  This
Monte Carlo
time series was Fourier transformed in the same manner as the
data were.  The $\chi^{2}$ of the Monte
Carlo PDS to the data PDS was
calculated for frequencies above 100~Hz.  This procedure was 
repeated for a grid of parameter values whose origin was defined
by the initial fit to the $\Delta t_{\gamma}$ distribution.  The
resulting best fit PDS defined the effective Poisson noise floor.
The
resultant noise corrected PDS for the HEAO data is shown in
Fig.~\ref{hcygpwr2.eps}. Fig.~\ref{hcygpwr3.eps} shows the region
of the PDS above 10 Hz to better examine the power at high
frequencies. No statistically significant power above the noise
floor is observed above 25 Hz. This is consistent with our
assumption that power above 100 Hz is attributable to Poisson
noise. \placefigure{hcygpwr2.eps} \placefigure{hcygpwr3.eps}

\subsection{Relative integral power analysis}

M84 derived a statistic, that they called the relative integral power,
to quantify aperiodic variability. The details of their approach are
described in Section III of their paper.
 
The new statistic, $P_{rel}$, of M84 is defined as the total discrete
Fourier transform power of the mean subtracted time series divided by
the square of the total number of x-ray counts, $N^{2}$, in the time series
of length $T$ divided into $m$ equal length bins.
\begin{equation}
P_{rel} \equiv{1 \over {N^2}} ~ {\sum_{j=-{{m\over 2} -1}}^{m
\over 2}}{(\vert a_{j} \vert^{2} - a_{0}^{2})} =
\frac{\chi^{2}}{N}
\label{relpow}
\end{equation}
where the $a_{j}$ are the standard Fourier coefficients
\begin{equation}
a_{j} = \sum_{k=0}^{m-1}x_{k}e^{2\pi ijk/m},
\label{ajdef.eq}
\end{equation}
$x_{k}$ is the number of
events in the $k$th time bin, and $a_{j}$ is the Fourier coefficient at
frequency  $f_{j}(= j/T)$.

The distribution of power variability over all possible frequencies
forms the Fourier power spectrum. With the Leahy normalization, this
power spectrum is given by Leahy~\it et~al.\rm (1983);
\begin{equation}
P_{j} = {2 \vert a_{j} \vert^{2} \over N},  j = 1,..., m/2-1
\end{equation}
where $N$ is the total number of x-ray counts observed in the time
interval 0$\rightarrow$T.

The M84 analysis did not include corrections for dead time or
instrumental effects. We have derived an approximation to the relative
integral power that allows for simple corrections to equation 1 of M84 for
these effects. We define $\chi^{2}$ as,
\begin{equation}
\chi^{2} = \sum_{j=1}^{{m \over 2} -1} P_{j} + \frac{1}{2}P_{m
\over 2} \label{leahyrel.eq}
\end{equation}
where $m$ is the number of time bins in the data segment under
consideration.  As in M84, the 9 minutes of HEAO A-1 data is divided
into $L$ contiguous data segments of width $\Delta t_{seg}$, 
each containing $m$ (=
10) time bins, and various quantities were calculated. This is
repeated with $T\equiv \Delta t_{seg}$ = 0.3 ms, 1 ms, 3 ms, 10 ms, 
30 ms and 100 ms.

The average of the $\chi^{2}$ in Equation~\ref{leahyrel.eq} 
over the entire ensemble 
of 10 bin
data segments for a given $\Delta t_{seg}$  can be approximated by
\begin{equation}
\langle \chi^{2} \rangle \approx {m-1 \over 2} \langle P \rangle
\label{chiens.eq}
\end{equation}
where $\langle P \rangle$ 
is the average Leahy-normalized power over the entire
ensemble of 10 bin data segments ($L \approx$ 9 minutes/$\Delta t_{seg}$) 
and
the set of  frequencies 
$f_{j} = 1/\Delta t_{seg} , 2/\Delta t_{seg}, ...
m/2 \Delta t_{seg}$.
Using equation~\ref{chiens.eq} for the average 
$\chi^{2}$ in equation 1 of 
M84 yields
\begin{equation}
\langle P_{rel} \rangle \approx {{{ m-1} \over 2}{[\langle P
\rangle - \langle P \rangle_{noise}]} \over {(\langle N \rangle
-1)}} \label{relleahy.eq}
\end{equation}
for each $\Delta t_{seg}$. The expected noise is
simply proportional to the Poisson floor.  

\begin{figure}
\plotone{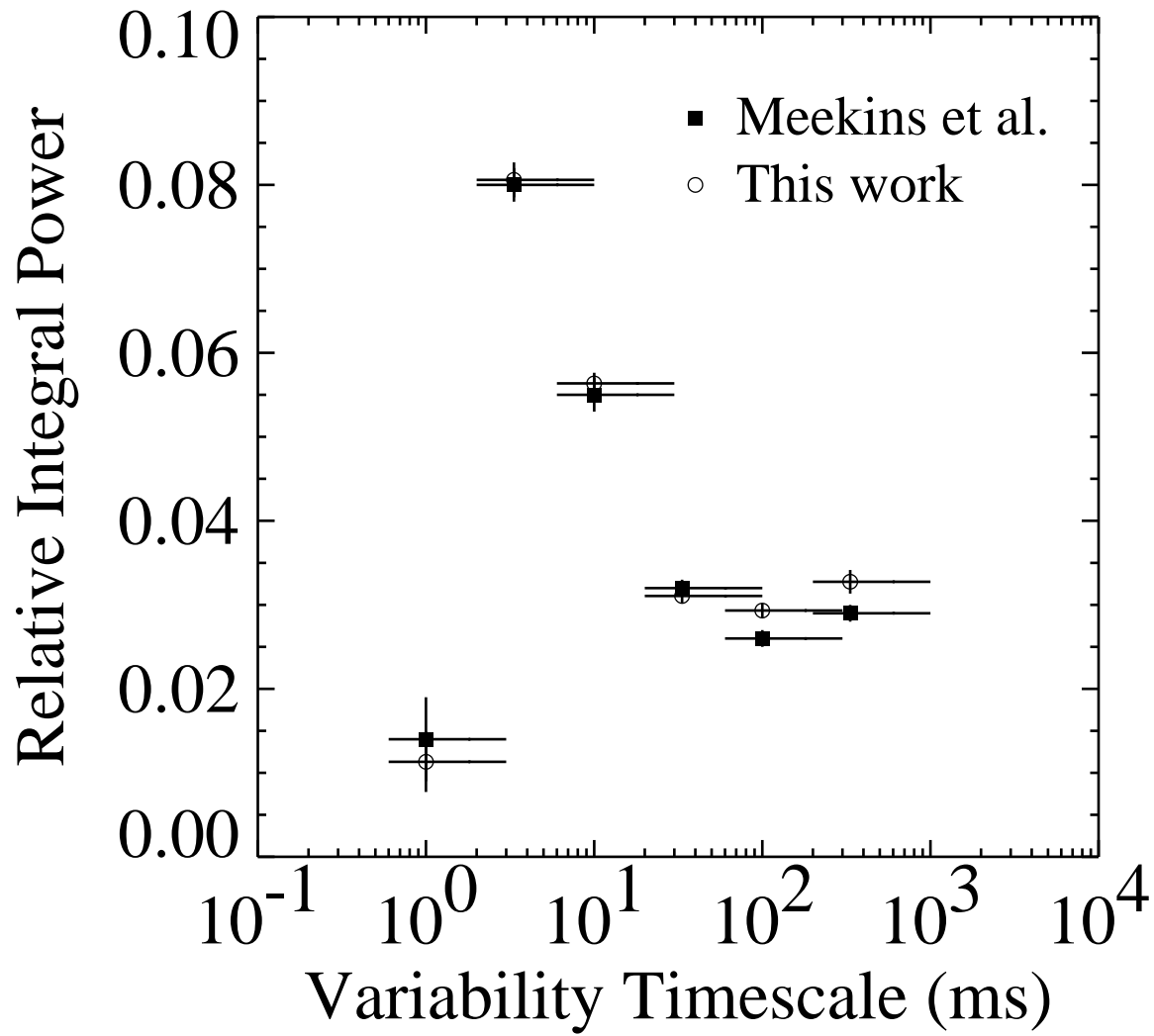} \caption{Reanalysis of the HEAO A-1
Cygnus X-1 data using the relative integral power method of M84.
Shown here are both the original M84 results and our reanalysis using
the same method, as discussed in the text.} \label{meekold.eps}
\end{figure}

The M84 analysis of the
HEAO A-1 HBR observation of Cyg~X-1 found an excess of variability at
time scales above 1 ms and a sharp cutoff at about 1 ms . We have
reanalyzed the same observation using their method without dead time
and instrumental corrections and have found excellent consistency with
their results. There are some minor discrepancies which can be
attributed to differences in the bin offsets and the use of  a different
digitization of the original analog tape. The comparison of our
analysis to that of M84 is shown in Figure~\ref{meekold.eps}.  
\placefigure{meekold.eps}

\begin{figure}
\plotone{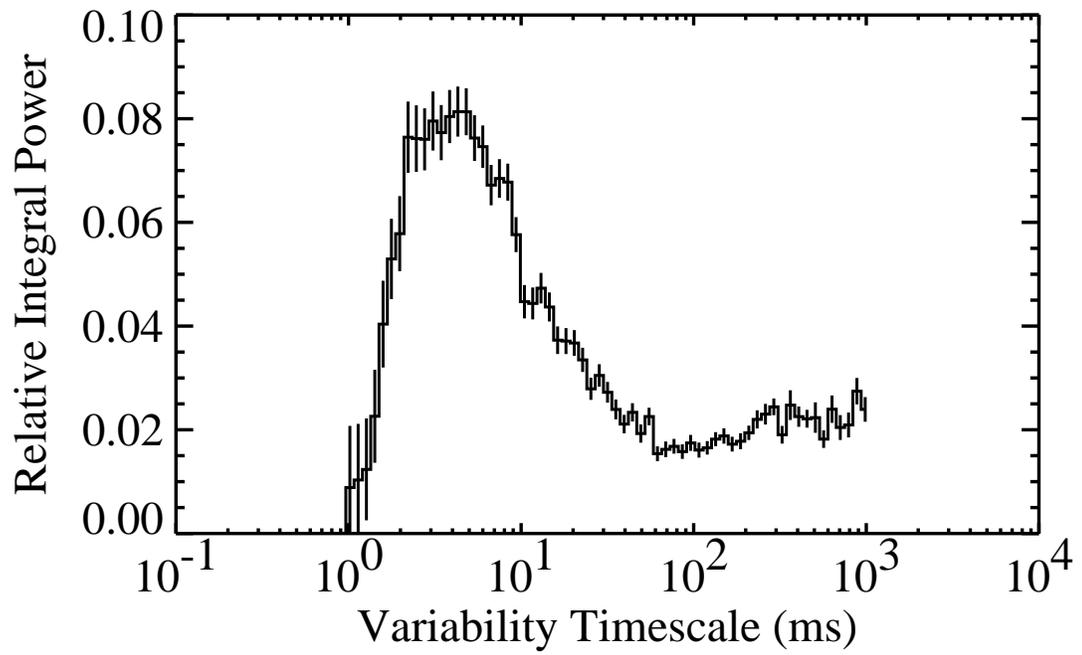} \caption{The HEAO A-1
relative integral power with no corrections
applied.}
\label{meekprecorr.eps}
\end{figure}

\begin{figure}
\plotone{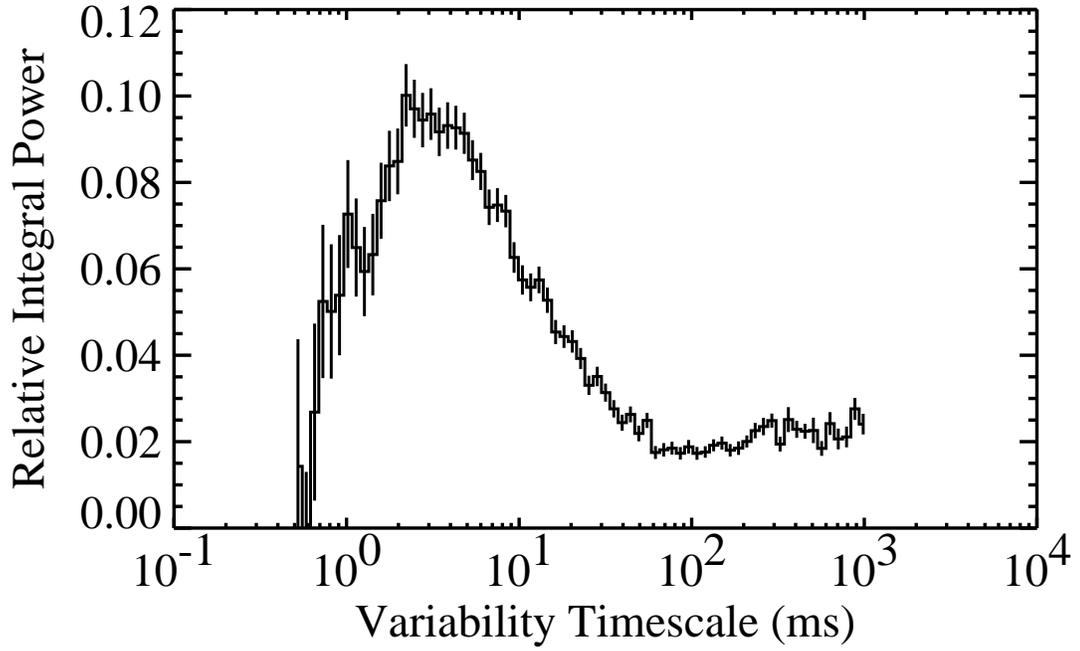} \caption{The HEAO A-1 
Poisson dead time corrected relative integral power.  The
corrections applied assume that the underlying difference in photon
arrive times ($\Delta t_{\gamma}$)
distribution is a simple offset exponential distribution.}
\label{meekdead.eps}
\end{figure}

\begin{figure}
\plotone{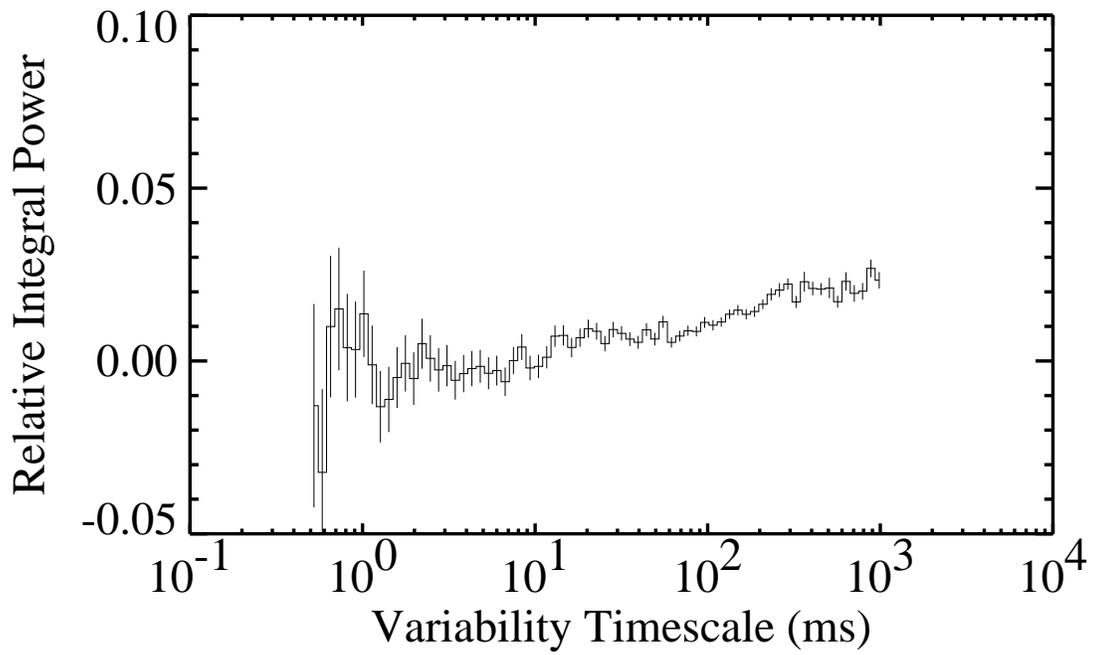} \caption{The HEAO A-1 
relative integral power corrected dead time instrumental effect.
The corrections applied assume
Equation~\ref{hyper} describes the underlying diffence in
photon arrival times ($\Delta t_{\gamma}$)
distribution.} \label{meekcorr.eps}
\end{figure}

By ignoring
instrumental effects, M84 chose a value of 2 for the noise floor,
where 2 is the value at all frequencies of the Fourier transform of
a Poisson source in the Leahy normalization. Using
Equation~\ref{hyper} as the probability distribution for the noise
floor, we have calculated the expected noise floor in
Equation~\ref{relleahy.eq}.   We binned the data in a different manner
than the original M84 analysis.  The uncorrected results with the new
binning are shown in Figure~\ref{meekprecorr.eps}.  The  shape and
normalization are in good agreement with that of the original M84
work, shown in Figure~\ref{meekold.eps}.
\placefigure{meekprecorr.eps}
Figure~\ref{meekdead.eps} shows the results of correcting
for standard Poisson dead time.  Note that the normalization
is increased and that the peak is broader, enhancing the
effect observed by M84.
\placefigure{meekdead.eps}
The results of our analysis, which has been corrected for
dead time and instrumental effects, 
of the HEAO A-1 observation of Cyg~X-1 are
shown in Figure~\ref{meekcorr.eps}.  We see no evidence for the
previously reported rise in the relative integral power, once the
corrections for the previously uncorrected instrumental effects 
or the manifestation of the known reset problem are
applied. \placefigure{meekcorr.eps}

\subsection{Power spectrum Analysis of RXTE Data}

\begin{figure}
\plotone{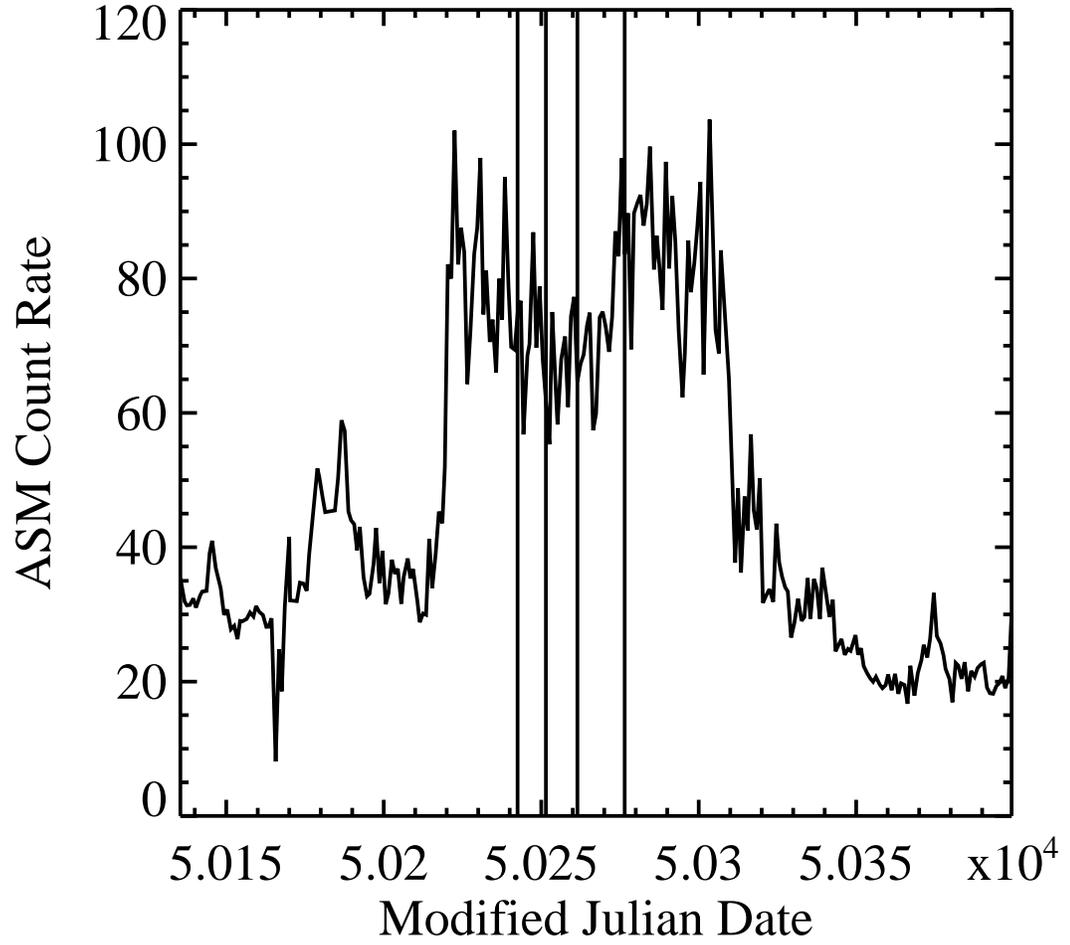}
\caption{The RXTE All Sky Monitor light curve for Cyg~X-1.  The vertical
lines represent the dates of our observations.}
\label{asm.eps}
\end{figure}

We have analyzed four RXTE/PCA observations of Cyg~X-1, see Table
1. 
The RXTE data were recorded with 4~$\mu$s
time resolution. 
During all four observations the source was in the high (soft)
state. Figure~\ref{asm.eps} shows the RXTE All Sky Monitor (ASM)
light curve for Cyg~X-1 around the time of our observations.
\placefigure{asm.eps}

\begin{figure}
\plotone{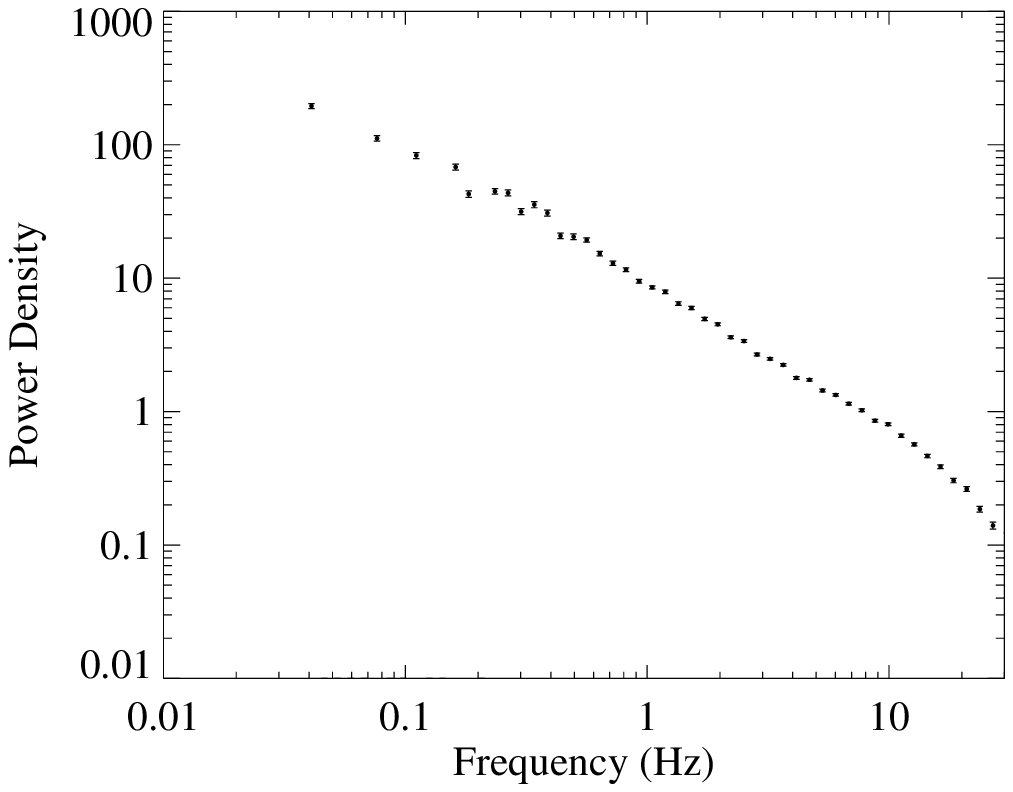}
\caption{The Poisson noise subtracted PDS for the RXTE observations of
Cyg~X-1}
\label{xtepds.ps}
\end{figure}

\begin{figure}
\plotone{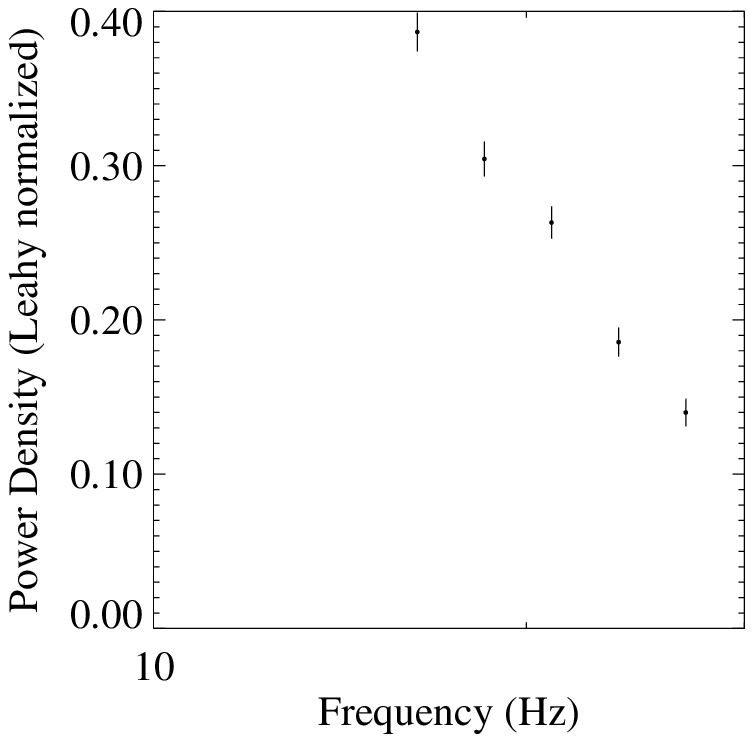}
\caption{Blow up of the
Poisson noise subtracted PDS for the RXTE observations of Cyg~X-1.}
\label{xtepdsexp.eps}
\end{figure}

We binned the data into 50 microsecond bins. The light curves were
divided into equal segments of 26 seconds. An FFT was performed on
each data segment. The results were averaged over all segments and
over equal logarithmic frequency intervals. We used the Leahy
normalization for the PDS. The dead time corrected
Poisson noise power
was then subtracted from the PDS obtained to
yield the remaining signal above the noise. To determine the
Poisson noise floor, we calculated the Poisson power spectrum,
correcting for nonparalyzable dead time using Equation~44 in Zhang
\it et~al. \rm (1995) with a dead time of 10 microseconds
(\cite{zhang98}). Corrections were not made for the energy
dependent dead time or for very large events. However, below 30~Hz,
these corrections are not significant and can be ignored
(\cite{cui97}).  Figure~\ref{xtepds.ps} shows the Poisson 
noise subtracted PDS for these observations.  Figure~\ref{xtepdsexp.eps}
shows the 10--30~Hz region on a linear scale to show the
behavior of the PDS as it approaches the limit imposed by the
corrections.  
\placefigure{xtepds.ps}

\placefigure{xtepdsexp.eps}

\begin{figure}
\plotone{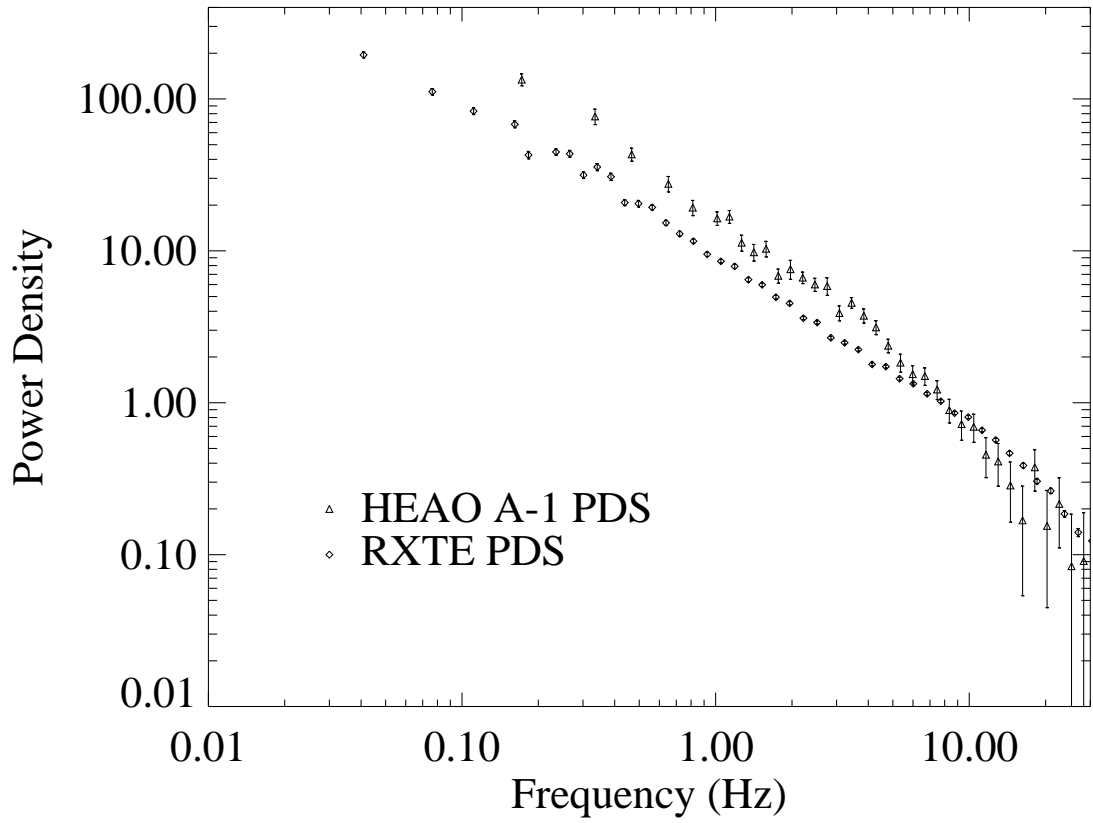}
\caption{The PDS's for both HEAO A-1 and RXTE.  For
comparison we show the PDS's for both instruments using
the Leahy normalization.}
\label{heaoxtepds.eps}
\end{figure}

Figure~\ref{heaoxtepds.eps} shows the PDS's for both RXTE and HEAO~A-1
on the same axes.  

\section{Results}

\subsection{HEAO A-1}

As reported above, we have discovered either an unknown
instrumental effect or a manifestation of the known reset problem in
the HEAO A-1 high bit rate data. This effect could not have been
discovered using the binned 5 ms and 320 ms
timing resolution modes of HEAO A-1 because their Nyquist
frequencies are too low.  
The effect is observable only with the higher 8~$\mu$s time resolution
of the HBR data or it is a mode-dependent problem that only occurs in
the HBR data mode.
Once this effect is taken into account, we observe no
evidence for the rise or the sharp cutoff in the relative integral
power for Cyg~X-1 reported by M84.

We observe excess power with a 95\% confidence level at
frequencies below 25~Hz in the noise subtracted PDS from Cyg
X-1 in its hard state.  Above 30--40~Hz, the noise subtracted PDS is
consistent with the null hypothesis. In the region where excess
power is significant, we find that the spectral shape can be
described by a power law spectrum with a break in the spectrum at
3~Hz.  From 0.1 to 3~Hz the spectral index is $1.20\pm0.08$ and
above the 3~Hz the spectrum steepens to $1.7\pm0.2$.

\subsection{RXTE}

Our results are consistent with previously
published results (\cite{cui97,bell96}).  
Below 30~Hz we find that the spectral shape can
be described by a broken power law with the break 
occurring at about 10~Hz.  Below
10~Hz the spectral index is 1.05$\pm$0.01 and between 10 and 30~Hz
the spectral index steepens to 1.75$\pm$0.03.

The lack
of corrections for very large events and energy dependent dead time
in the standard RXTE corrections
make it impossible to extend the search for excess power beyond
about 30~Hz at this time.  
There is adequate data in the sample to extend the
search to higher frequencies once these additional corrections are
developed.
Additional work on these corrections is necessary
to exploit the full timing capabilities of RXTE to search for excess
power using this method.

\section{Conclusions}

In light of our discovery of either a new instrumental effect 
or a correction for the known reset problem that accounts
for the observed relative integral power of M84 there is no longer
any evidence for model-independent aperiodic variability on
millisecond timescales from Cyg~X-1.  This lack of observed
variability does not rule out its existence. RXTE should have the
capability to make measurements of excess power at millisecond
time scales once the appropriate corrections are available.

Our results for the Cyg~X-1 PDS's are consistent with previous 
measurements in the hard and soft states.
Previous measurements of Cyg~X-1 often show a flat PDS below 
about 0.1~Hz, although the location of the maximum frequency
varies from about 0.04 to 0.4~Hz.  The minimum frequency 
studied here is only
0.1~Hz, and we see no indication of a flat PDS in our data.  
Our RXTE spectral shape determination is consistent with other
RXTE observations (\cite{cui97,bell96}) made around the
same time.  In both sets of observations the spectral shape is
similar to that observed for most black-hole candidates in
the same state (\cite{vdk95}).

In order to extend the range of the types of searches into the regime
where they can start to impact the models discussed above, we must
make several improvements to the data and the techniques.  At present,
the limiting factor the RXTE data is the lack of adequate background
subtraction at high frequencies.  This is being worked on (\cite{zhang98})
but is not available.  Cross checks on Poisson sources, as presented
here using Cas A, are invaluable in searching for uncorrected dead time
and instrument difficulties.  To maximize the utility of the cross checks,
the cross checking
observations are best done in the same data taking modes and at around
the same 
times as the the observations of the source under study.  Unfortunately,
the current available
instrument data sets do not have these properties.

\acknowledgments

Work supported by Department of Energy contract DE-AC03-76SF00515, 
NASA RXTE Guest Investigator Grand, the Office of Naval Research, 
and Stanford University.  

\clearpage

\begin{deluxetable}{cccccc}
\tablecaption{HEAO A-1 and RXTE Observations \label{observ.tbl}}
\tablehead{
\colhead{Instrument} &  \colhead{Source} &
\colhead{Date}      &   \colhead{Obs. Time (UT)} &
\colhead{Time res. ($\mu$s)} & \colhead{Time (s)}
}
\startdata
HEAO A-1 &  Cyg~X-1 & 1978 May 07  &       N/A           &    8	&  510   \nl
HEAO A-1 &  Cas A   & 1978 Aug 02  &       N/A           &    8	&  540   \nl
RXTE PCA &  Cyg~X-1 & 1996 Jun 08  &   03:14 -- 03:36    &    4 &  1362  \nl
RXTE PCA &  Cyg~X-1 & 1996 Jun 17  &   03:19 -- 04:03    &    4 & 876   \nl
RXTE PCA &  Cyg~X-1 & 1996 Jun 27  &   05:14 -- 05:47    &    4 & 858   \nl
RXTE PCA &  Cyg~X-1 & 1996 Jul 12  &   12:31 -- 12:59    &    4 & 726   \nl
\enddata
\end{deluxetable}


\clearpage

\clearpage
\begin{thebibliography}{}
\bibitem[Bao \& $\O$stgaard 1995]{bao95} Bao, G. and $\O$stgaard, E., 1995
\apj, 443, 54
\bibitem[Belloni \it et~al. \rm 1996]{bell96} Belloni, T., \it{et~al.}\rm, 1996,
\apj, 472, L107
\bibitem[Chakrabarti \& Titarchuck 1996]{chak96} Chakrabarti, S. K. and
Titarchuk, L. G., 1996, \apj, 455, 623
\bibitem[Cui \it et~al. \rm 1997]{cui97} Cui, W., Zhang, S. N., Focke, W.,
and Swank, J. H., 1997, \apj, 484, 383
\bibitem[Eadie \it et~al.\rm 1971]{eadie71} Eadie, W. T., Drijard, D.,
James, F. E., Roos, M. and Sadoulet, B., 1971, Statistical Methods 
in Experimental Physics, (New York:North-Holland)
\bibitem[Leahy \it et~al. \rm 1983]{leahy} Leahy, D. A., \it et~al. \rm,
1983, \apj, 266, 160
\bibitem[Liang 1998]{liang98} Liang, E. P., 1998, Phys. Rep., 302, 67
\bibitem[Lochner, Swank \& Szymokowiak 1989]{loch89} Lochner, J. C.,
Swank, J. H., and Szymokowiak, A. E., 1989, \apj, 337, 823
\bibitem[Lochner, Swank \& Szymokowiak]{loch91} Lochner, J. C.,
Swank, J. H., and Szymokowiak, A. E., 1991, \apj, 376, 295
\bibitem[Meekins \it et~al. \rm 1984]{meek84} Meekins, J. F., \it el al.\rm,
1984, \apj, 278, 288
\bibitem[Negoro \it et~al. \rm 1995]{nego95} Negoro, H., Kitamoto, S.,
Takeuchi, M., and Mineshige, S., 1995, \apj, 452, L49
\bibitem[Nowak \& Wagner 1995]{nowak95} Nowak, M. A. and Wagoner, R. V.,
1995, \mnras, 274, 37
\bibitem[Narayan 1996]{nara96} Narayan, R., 1996, \apj, 462, 136
\bibitem[Perez \it et~al. \rm 1997]{perez97} Perez, C. A., Silbergleit, A. S.,
Wagoner, R. V., and Lehr, D. E., 1997, \apj, 476, 589
\bibitem[Press \& Schechter 1974]{press74} Press, W. and Schechter, P.,
1974, \apj, 193, 437
\bibitem[Rothschild \it et~al. \rm 1974]{roth74} Rothschild, R. E., Boldt, E. A.,
Holt, S. S., and Serlemitsos, P. J., 1974, \apj, 189, L13
\bibitem[van der Klis 1995]{vdk95} van der Klis, M. 1995, X-Ray Binaries,
1995, 252
\bibitem[van der Klis 1997]{vdk97} van der Klis, M. 1995, astro-ph/9710016
\bibitem[Wallinder \it et~al. \rm 1992]{wall92} Wallinder, F. H., Kato, S., and
Abramowicz, M. A., 1992, \aap, 4, 79
\bibitem[Weisskopf \& Sotherland 1978]{weiss78} Weisskopf, M. C. and
Sutherland, P. G., 1978, \apj, 221, 228
\bibitem[Wood \it et al. \rm 1984]{wood84} Wood, K. S.\it el al.\rm,
1984, \apjs, 56, 507  
\bibitem[Zhang et~al. 1995]{zhang95} Zhang, W., \it el al.\rm,
1995, \apj, 449, 930
\bibitem[Zhang \& Jahoda 1998]{zhang98} Zhang, W. and Jahoda, K., 1998,
in preparation
\end{thebibliography}
\end{document}